\documentclass[prx,twocolumn,aps,superscriptaddress]{revtex4-1}
\usepackage{graphicx}
\usepackage{amsfonts}
\usepackage{amsmath,amssymb,amsthm,bm,multirow,longtable}
\usepackage{epsfig}
\usepackage{dcolumn}
\usepackage{epstopdf}
\usepackage{comment}
\usepackage{color}
\usepackage{hyperref}
\usepackage{bm}
\usepackage{float}

\begin{document} 

\title{Geometric formalism for constructing arbitrary single-qubit dynamically corrected gates}

\author{Junkai Zeng}
\affiliation{
 Department of Physics, Virginia Tech, Blacksburg, Virginia 24061, USA
}
\author{C. H. Yang}
\affiliation{
 Centre for Quantum Computation and Communication Technology, School of Electrical Engineering and Telecommunications, The University of New South Wales, Sydney, NSW 2052, Australia
}
\author{A. S. Dzurak}
\affiliation{
 Centre for Quantum Computation and Communication Technology, School of Electrical Engineering and Telecommunications, The University of New South Wales, Sydney, NSW 2052, Australia
}
\author{Edwin Barnes}
 \email{efbarnes@vt.edu}
 \affiliation{
 Department of Physics, Virginia Tech, Blacksburg, Virginia 24061, USA
}
\begin{abstract}
Implementing high-fidelity quantum control and reducing the effect of the coupling between a quantum system and its environment is a major challenge in developing quantum information technologies. Here, we show that there exists a geometrical structure hidden within the time-dependent Schr\"odinger equation that provides a simple way to view the entire solution space of pulses that suppress noise errors in a system's evolution. In this framework, any single-qubit gate that is robust against quasistatic noise to first order corresponds to a closed three-dimensional space curve, where the driving fields that implement the robust gate can be read off from the curvature and torsion of the space curve. Gates that are robust to second order are in one-to-one correspondence with closed curves whose projections onto three mutually orthogonal planes each enclose a vanishing net area. We use this formalism to derive new examples of dynamically corrected gates generated from smooth pulses. We also show how it can be employed to analyze the noise-cancellation properties of pulses generated from numerical algorithms such as GRAPE. A similar geometrical framework exists for quantum systems of arbitrary Hilbert space dimension.
\end{abstract}

\maketitle
In recent years, novel information technologies based on the principles of quantum mechanics have attracted growing interest from both academia and industry. For example, a quantum computer could enable us to tackle certain problems exponentially faster than an ordinary classical computer \cite{Nielsen_Chuang}. For decades, people have been striving to overcome one of the main obstacles to realizing this and other proposed quantum technologies, namely the decoherence caused by the coupling between a qubit and its noisy environment \cite{chirolli2008decoherence,Bialczak_PRL07}. Quantum error correcting codes provide a way to surmount this problem; however, it remains a challenging task to raise the fidelity of qubit control above the error thresholds that determine when these codes work \cite{gottesman2010introduction,preskill1998reliable,aliferis2007level}, although considerable experimental progress in recent years has brought this within reach in a number of physical systems \cite{Veldhorst_Nature2015,Ballance_PRL2016,Takita_PRL2017,Nichol_npjQI2017,Watson_Nature2018,Zajac_Science2018,barends2014superconducting,yoneda2018quantum}. 

Inspired by the Hahn spin echo pulse introduced in the context of nuclear magnetic resonance \cite{Hahn_PR50}, a wide range of techniques for implementing dynamical decoupling (DD), or more generally dynamically corrected gates (DCGs) \cite{khodjasteh2010arbitrarily}, have been developed in which deviations in a system's evolution caused by noise fluctuations or parameter inhomogeneities can be corrected by applying carefully designed driving pulses. Early work in DD mainly made use of instantaneous $\pi$ pulses ($\delta$-pulses) to flip the qubit state one or more times during the evolution such that the coupling to the environment is effectively undone \cite{Carr_Purcell,Meiboom_Gill,Viola_PRA98,Uhrig_PRL07,yang2008universality}. In the context of both DD and DCGs, methods based on square pulses have also been developed \cite{Wang_NatComm12,khodjasteh2012automated,Wang_PRA14,Merrill_Wiley14}. However, in a real experiment $\delta$-function or square waveforms can only be generated approximately since they would require infinite power or arbitrarily fast electronics to realize exactly. This leads to an imperfect cancellation of errors and thus diminishes the performance of such DCG schemes, especially in systems that evolve on nanosecond time scales. Many DCG schemes are based on concatenating two or more noisy quantum operations that together produce the desired gate while their errors cancel up to some order \cite{viola2003robust,khodjasteh2009dynamically,khodjasteh2009dynamical,khodjasteh2010arbitrarily,khodjasteh2012automated,Kestner_PRL13,Merrill_Wiley14,braun2014concurrently,CalderonVargas_PRL2017,Gungordu_PRB2018,Buterakos_PRB2018a,Buterakos_PRB2018b,Gungordu_PRB2018}. While some of these protocols can work for any choice of the pulse shapes used in the sequences, they leave open the possibility that more efficient methods based on the application of single, shaped pulses exist. 

Searching for control pulse waveforms that implement DCGs in a single shot is difficult (aside from a few simple cases like $\delta$ or square pulses) because the time-dependent Schr\"odinger equation cannot be solved analytically in general, even for a two-level system. Numerical methods  have been shown to be quite effective in many cases \cite{palao2002quantum,brif2010control,yang2018optimization}, but using these to find globally optimal waveforms that respect the constraints of a given physical system can be challenging, although there has been some recent progress in this direction \cite{glaser2015training,suter2016colloquium,doria2011optimal,caneva2011chopped}. Analytical methods should really be viewed as complementary to numerical techniques, where they can provide additional insight into why such techniques work or provide starting pulses that can speed up numerical algorithms. Analytical approaches that have been developed to circumvent the insolubility of the Schr\"odinger equation include methods based on Chebyshev polynomial approximations or reverse-engineering techniques  \cite{fanchini2007continuously,jones2012dynamical,Barnes_PRL12,Barnes_SciRep15}, however, these approaches have not yet succeeded in providing pulses that implement arbitrary DCGs, either because the methods only produce robust identity operations by design or because the constraint equations that determine the pulses are too difficult to solve. 

We recently introduced an alternative analytical approach that works for a U(1) subset of single-qubit DCGs that is remarkably simple to use \cite{zeng2018general}. The U(1) subset is comprised of rotations about an axis orthogonal to the noise term in the qubit Hamiltonian. We showed that all pulses which generate such rotations while canceling the noise to first order correspond to the set of all closed curves lying in a two-dimensional plane. The simplicity of the method lies in the fact that the pulse waveforms are precisely given by the curvature of these curves, a quantity which is very easy to compute. Moreover, we showed that plane curves that enclose zero net area yield pulses that cancel noise up to second order. Follow-up works showed that this method enables one to find the fastest possible pulses that implement a desired DCG within this U(1) subset \cite{zeng2018fastest}, and that it can be extended to suppress not only noise transverse to the pulse but also noise in the pulse amplitude \cite{robert2018conditions}. However, the fact that this method is restricted to a particular U(1) subset means that it cannot be used to generate the robust universal gate set needed for most quantum information applications.

In this work, we show that there exists a geometrical structure hidden within the time-dependent Schr\"odinger equation that provides a simple way to identify all pulse waveforms that implement DCGs spanning the entire SU(2) space of single-qubit operations subject to quasistatic noise. We show that any closed three-dimensional space curve corresponds to a qubit evolution operator in which the leading-order error vanishes, and that the driving fields which implement this evolution can be read off from the curve's curvature and torsion. A relation between first-order robust evolution and closed curves is expected based on general Lie-algebraic considerations \cite{Merrill_Wiley14}, however an explicit protocol that yields all DCGs using this perspective has been lacking. Furthermore, we show that all pulses which implement dynamical gate correction up to second order are in one-to-one correspondence with closed curves whose projections onto three mutually orthogonal planes each enclose zero net area. We provide explicit examples to demonstrate how the method works. We also briefly describe how a similar framework holds for higher-dimensional Hilbert spaces.

A driven qubit subject to a single source of quasistatic noise can generally be described by the Hamiltonian
\begin{equation}
\begin{split}
\mathcal{H}(t)&=\mathcal{H}_0(t)+\delta\mathcal{H}\\
&=\frac{\Omega(t)\cos\Phi(t)}{2}\sigma_x+\frac{\Omega(t)\sin\Phi(t)}{2}\sigma_y+\delta\beta\sigma_z,
\end{split}
\label{eq:hamil}
\end{equation}
where $\sigma_x$, $\sigma_y$ and $\sigma_z$ are Pauli matrices, and $\Omega(t)$ and $\Phi(t)$ determine two driving fields that are applied along orthogonal directions. $\delta\mathcal{H}$ is the quasistatic noise term, which we assume is weak compared to the driving fields: $\|\delta\mathcal{H}\|\ll \frac{1}{T}\int_0^T \Omega(t)dt$, where $T$ is the duration of the gate. We also assume that the noise is slow compared to the pulse duration so that $\delta\beta$ is treated as a constant (but unknown) fluctuation parameter. While the noise term may lie along any direction depending on the type of system, it is sufficient to only consider the case where the noise is transverse to the driving, as in Eq.~\eqref{eq:hamil}, 
because we can always transform to a frame in which the Hamiltonian takes this form. For example, for a Hamiltonian given by $\tilde{\mathcal{H}}(t)=\frac{\tilde{\Omega}_x(t)}{2}\sigma_x+\frac{\tilde{\Omega}_z(t)}{2}\sigma_z+\delta\beta\sigma_z$, the transformation operator that does this is $R(t)=\text{diag}\{e^{\frac{i}{2} \int_0^t\tilde{\Omega}_z(\tau)dt},e^{-\frac{i}{2} \int_0^t\tilde{\Omega}_z(\tau)dt}\}$. Also note that if we start with a Hamiltonian with driving along all three axes, this can again be transformed into Eq.~\eqref{eq:hamil} using a similar transformation operator.

It is convenient to transform the Hamiltonian into the interaction picture, where we have 
\begin{equation}
	\mathcal{H_I}(t)=U_0^\dagger(t)\sigma_zU_0(t)\delta\beta,\label{eq:inthamil}
\end{equation} 
where $U_0(t)$ is the evolution operator associated with the original error-free Hamiltonian $\mathcal{H}_0(t)$, which can be generically parameterized as
\begin{equation}
\begin{split}
 	U_0(t)&=\left(
\begin{array}{cc}
 u_1(t) & -u_2^*(t)\\
 u_2(t) & u_1^*(t)\\
\end{array}
\right),\\
u_1(t)&=e^{\frac{1}{2} i (\theta (t)+\phi (t))} \cos \left(\frac{\chi (t)}{2}\right),\\
u_2(t)&=-i e^{\frac{1}{2} i (\phi (t)-\theta (t))} \sin \left(\frac{\chi (t)}{2}\right).
\end{split}
\label{eq:evolution}
 \end{equation}
Requiring $U_0(0)=\mathbf{1}$ gives the initial conditions $\chi(0)=0$ and $\phi(0)=-\theta(0)$. Note that we cannot obtain an explicit analytical solution for even the error-free evolution, $U_0(t)$, in the case of arbitrary driving fields $\Omega(t)$ and $\Phi(t)$ because of the intractability of the time-dependent Schr\"odinger equation \cite{gangopadhyay2010,Barnes_PRL12}. Remarkably, this does not prevent us from obtaining the full solution space of DCGs for this problem, as we will see.

To obtain robust qubit operations, we need to require the evolution operator in the interaction picture to be the identity at the end of the evolution: $U_\mathcal{I}(T)=\mathbf{1}$. This in turn implies that the evolution in the lab frame will equal the target gate we want to perform. We can impose this constraint order by order using a Magnus expansion for $U_\mathcal{I}(T)$. The first two orders of the expansion involve the integrals
\begin{equation}
  	\begin{split}
  	&A_1(t)=\frac{1}{\delta\beta}\int_0^t\mathcal{H_I}(t_1)dt_1,\\
  	&A_2(t)=\frac{1}{2\delta\beta^2}\int_0^t dt_1\int_0^{t_1}dt_2\left[\mathcal{H_I}(t_1),\mathcal{H_I}(t_2)\right].
  	\label{eq:errcancellation}
  	\end{split}
  \end{equation}  
If we impose $A_1(T)=0$ and $A_2(T)=0$, then the first- and second-order errors in the evolution vanish, respectively. The problem is then to find the driving fields $\Omega(t)$ and $\Phi(t)$ that satisfy these conditions and thus generate DCGs. 

We tackle this problem by introducing the following geometrical framework. First decompose $A_1(t)$ into Pauli matrices: 
  \begin{equation}
  	A_1(t)=\mathbf{r}(t)\cdot\mathbf{\hat{\sigma}}=x(t)\sigma_x+y(t)\sigma_y+z(t)\sigma_z.\label{eq:A1}
  \end{equation}
Here, $\mathbf{r}(t)=(x(t),y(t),z(t))$ parameterizes a curve in three-dimensional Euclidean space that starts at the origin at time $t=0$: $\mathbf{r}(0)=(0,0,0)$. Noticing from Eqs.~\eqref{eq:inthamil} and \eqref{eq:errcancellation} that $[\dot{A_1}(t)]^2=\mathbf{1}$, it follows that $\|\dot{\mathbf{r}}(t)\|^2=1$, and thus $\mathbf{r}(t)$ is the natural arc-length parameterization of the curve. This can also be seen by plugging Eq.~\eqref{eq:evolution} into the definition of $A_1(t)$ from Eq.~\eqref{eq:errcancellation}, yielding
\begin{equation}
	\dot{\mathbf{r}}(t)=\left(-\sin \chi (t) \sin \phi (t),\sin \chi (t) \cos \phi (t),\cos \chi (t)\right),
	\label{eq:rdot}
\end{equation}
which is clearly a vector of unit length. Although we can parameterize a space curve in infinitely many ways, $\mathbf{r}(t)$ is special because for this parameterization, the length of the curve equals the evolution time.

It is clear from the definition of $\mathbf{r}(t)$, Eq.~\eqref{eq:A1}, that it measures the size of the first-order error in the evolution. We now show that it actually contains all the information about the evolution, not just the first-order error. To see this, first consider the second-order derivative of $A_1$:
   \begin{equation}
    	\begin{split}
    	\ddot{A_1}(t)=\ddot{\mathbf{r}}(t)\cdot\mathbf{\hat{\sigma}}= \frac{1}{ \delta \beta }\dot{\mathcal{H_I}}(t)=\frac{i}{\delta\beta} U_0^\dagger(t)[\mathcal{H}_0(t),\delta\mathcal{H}]U_0(t)
    	\end{split}\label{eq:ddotA1}
    \end{equation} 
Plugging Eq.~\ref{eq:hamil} into this result, we obtain
    \begin{equation}
    	\|\ddot{A_1}(t)\|_F=\|\ddot{\mathbf{r}}(t)\|=\frac{\|[\mathcal{H}_0(t),\delta\mathcal{H}]\|_F}{\delta\beta}=\Omega(t),
    \end{equation}
where $\|.\|_F$ is the Frobenius norm, scaled by the inverse of the square root of the dimension of the matrix. We have just shown that $\Omega(t)$ is precisely equal to the curvature of the curve, $\|\ddot{\mathbf{r}}(t)\|$, which is consistent with our earlier plane curve construction \cite{zeng2018general}. Thus, given a space curve, we can readily extract the corresponding driving field $\Omega(t)$ by computing the curvature. 

To see how we can obtain the rest of the control Hamiltonian, namely $\Phi(t)$, consider now the third-order derivative of $A_1(t)$. { Differentiating Eq.~\eqref{eq:ddotA1}, we obtain
\begin{equation}
\begin{split}
\dddot{A_1}(t)=\dddot{\mathbf{r}}(t)\cdot\hat\sigma=&-\frac{1}{\delta\beta}U_0^\dagger(t)\mathcal{H}_0(t)[\mathcal{H}_0(t),\delta\mathcal{H}]U_0(t)\\&+\frac{i}{\delta\beta}U_0^\dagger(t)[\dot{\mathcal{H}_0}(t),\delta\mathcal{H}]U_0(t)\\&+\frac{1}{\delta\beta}U_0^\dagger(t)[\mathcal{H}_0(t),\delta\mathcal{H}]\mathcal{H}_0(t)U_0(t).
\end{split}
\end{equation}
From this and the analogous expressions for $\dot{A_1}(t)$ and $\ddot{A_1}(t)$, it is straightforward to verify that the following formula holds: 
\begin{equation}
-2i\frac{\hbox{Tr}\{\dot{A_1}(t)\ddot{A_1}(t)\dddot{A_1}(t)\}}{\|[\dot{A_1}(t),\ddot{A_1}(t)] \|_F^2}=\dot\Phi.\label{eq:dotPhifromA1}
\end{equation}
Using the fact that $\hbox{Tr}\{\dot{A_1}(t)\ddot{A_1}(t)\dddot{A_1}(t)\}=\tfrac{1}{2}\hbox{Tr}\{[\dot{A_1}(t),\ddot{A_1}(t)]\dddot{A_1}(t)\}$, in combination with the Pauli operator identities $\hbox{Tr}\{(\mathbf{a}\cdot\hat\sigma)(\mathbf{b}\cdot\hat\sigma)\}=2\mathbf{a}\cdot\mathbf{b}$ and $[\mathbf{a}\cdot\hat\sigma,\mathbf{b}\cdot\hat\sigma]=2i(\mathbf{a}\times\mathbf{b})\cdot\hat\sigma$, we find that Eq.~\eqref{eq:dotPhifromA1} becomes the formula for the torsion $\tau(t)$ of the curve:
}
\begin{equation}
\tau(t)=\frac{\left(\dot{\mathbf{r}}(t)\times \ddot{\mathbf{r}}(t)\right)\cdot\dddot{\mathbf{r}}(t)}{\|\dot{\mathbf{r}}(t)\times \ddot{\mathbf{r}}(t)\|^2}=\dot\Phi(t).\label{eq:torsion}
\end{equation}
Therefore, we can obtain the full control Hamiltonian, ${\cal H}_0(t)$, from the space curve by computing its curvature and torsion. It should be noted that the integration constant we get by integrating Eq.~\eqref{eq:torsion} to obtain $\Phi(t)$ fixes the initial phases in the target evolution operator: $\mathcal{H}_0(t)=i \dot{U_0}(t)U_0^\dagger(t)\Rightarrow \Omega(0)e^{-i \Phi(0)}=e^{i \theta (0)} \dot\chi(0)\Rightarrow\Phi(0)=-\theta(0)=\phi(0)$. The key point here is that since we can extract ${\cal H}_0(t)$ from the space curve, it follows that the space curve determines the full qubit evolution, not just its leading-order error. This is essentially due to the fact that the Schr\"odinger equation for a two-level system is exactly an SU(2) representation of the Frenet-Serret equation for space curves \cite{lehto2015geometry}.

 \begin{figure}
 \centering
\includegraphics[width=0.7\columnwidth]{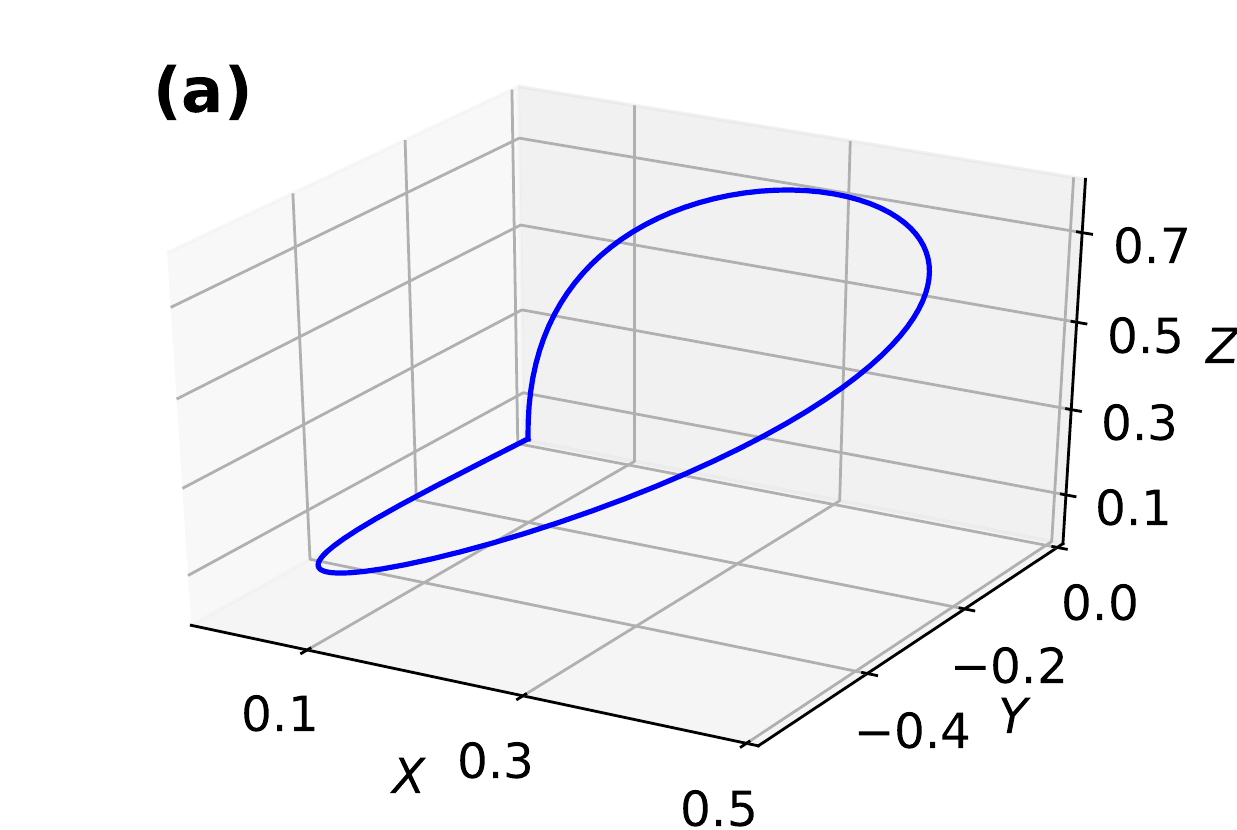}
\includegraphics[width=0.7\columnwidth]{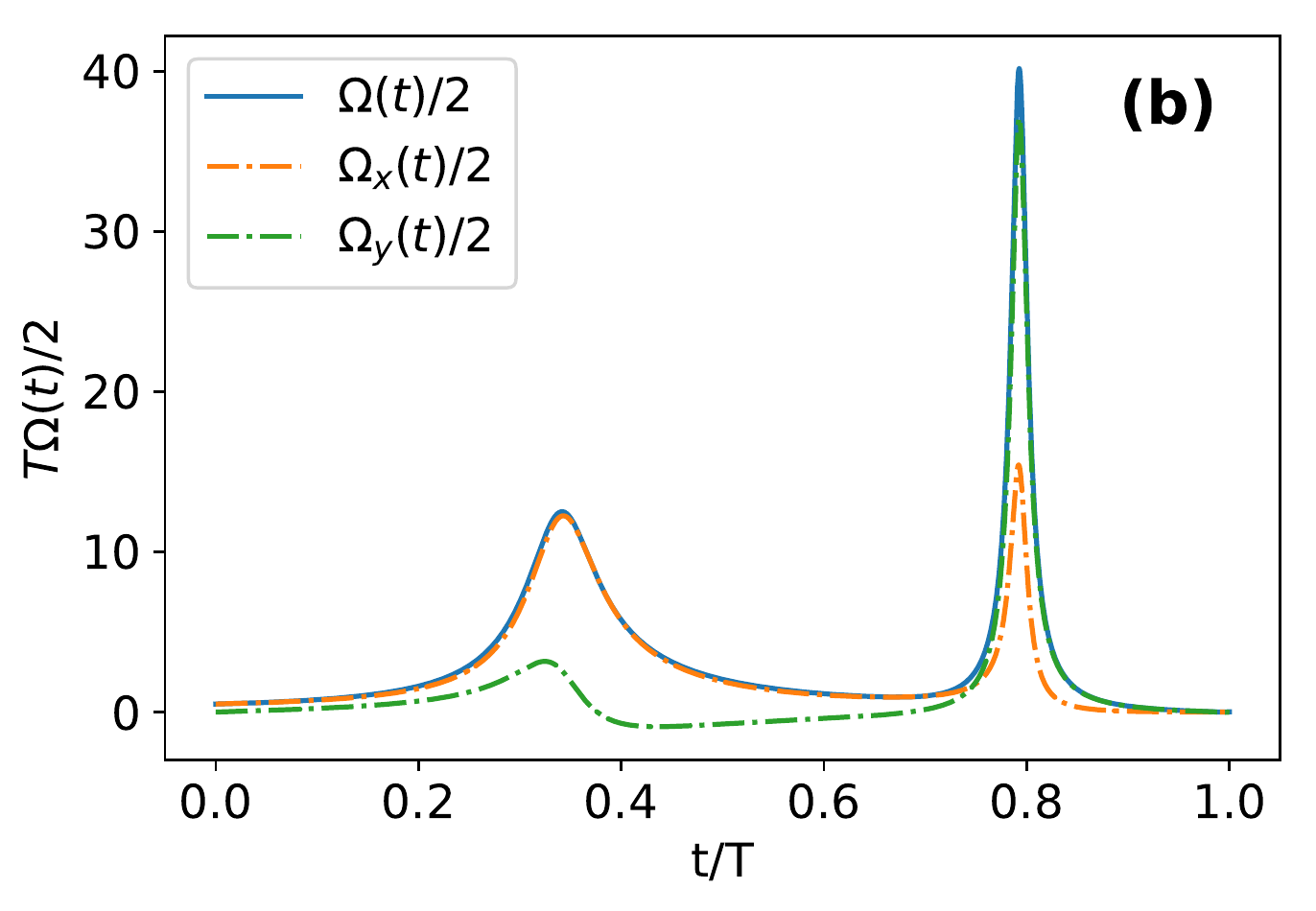}
\includegraphics[width=0.7\columnwidth]{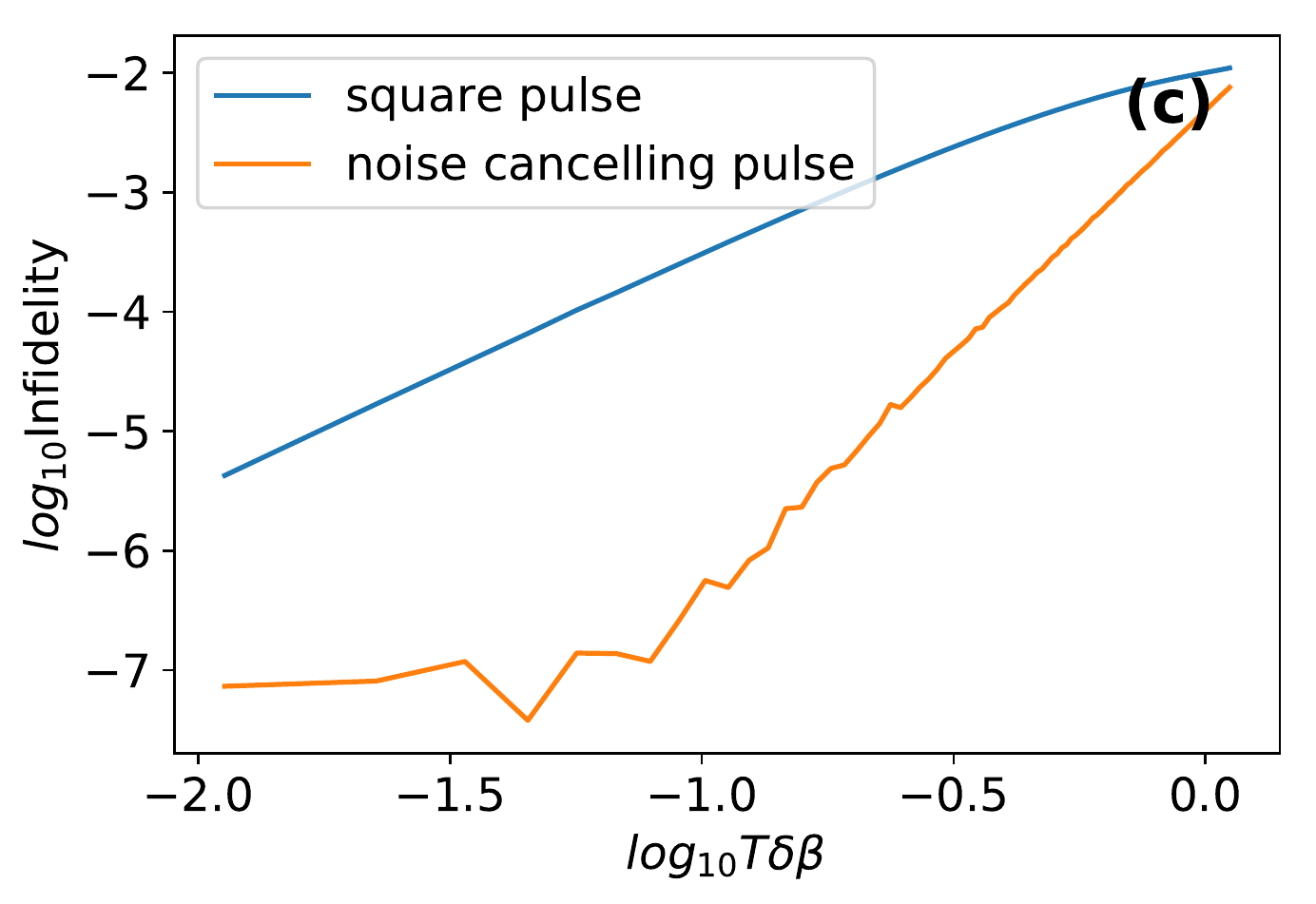}
\caption{Dynamically corrected Clifford gate $R(-\hat x+\hat y+\hat z,2\pi/3)$. (a) Closed space  curve. Here the curve is constructed as $r(l)=(1-l)r_1(l)+l r_2(l)$, where $r_1(l)=\sqrt{2} \sin (\pi l)\left(0,\sin ^2\left(\frac{\pi  l}{2}\right) , \cos ^2\left(\frac{\pi  l}{2}\right)\right)$ and $r_2(l)=\sqrt{2} \sin (\pi l)\left(\sin ^2\left(\frac{\pi  l}{2}\right) , \cos ^2\left(\frac{\pi  l}{2}\right),0\right)\cdot R_z(q)$, and where $l$ ranges from 0 to 1. Here $R_z(q)$ is the rotation matrix around $z$-axis for angle $q$. We have determined $q=1.6054$ numerically. { Changing $q$ rotates the curve rigidly about the $z$ axis but does not alter the relative orientation of the initial and final tangent vectors, and thus does not alter the target evolution.} (b) The pulse shape. Here, $\Omega_x=\Omega\cos\Phi$, $\Omega_y=\Omega\sin\Phi$. (c) Comparison of the log-log infidelity between the shaped pulse and naive square pulse.}
\label{fig:curve1}
\end{figure}

The fact that the space curves encode information about both the ideal evolution and the error is a powerful result in our effort to design DCGs. To ensure that the leading-order error vanishes at the end of the evolution, we simply impose $\mathbf{r}(T)=0$, i.e., the space curve must form a closed loop. Once we choose a closed curve, we can read off the control fields that perform the noise cancellation from its curvature and torsion. The only question that remains is whether we can simultaneously fix $U_0(T)$ to the desired target gate. Again, at first glance it would seem that one would need to solve the time-dependent Schr\"odinger equation to do this, however this is not necessary. From Eq.~\eqref{eq:rdot}, it is apparent that $\phi(T)$ and $\chi(T)$ are determined by the tangent vector of the curve at the final time, $\dot{\mathbf{r}}(T)$. The remaining angle in the target evolution can be determined from the total torsion (the integral of torsion along the curve): {
\begin{equation}
\begin{split}
    &\theta(T)-\theta(0)=\\&-\int_0^T \tau(t)dt-\arg \left[-i \ddot x(t) \dot y(t)+i \dot x(t) \ddot y(t)+\ddot z(t)\right] \bigg|^T_0.
\end{split}\label{eq:thetaFromTotalTorsion}
\end{equation}
This expression can be obtained by equating the arguments of the off-diagonal components of the matrices $\mathcal{H}_0(t)$ and $-i \dot{U_0}(t)U_0^\dagger(t)$. The resulting equation is seen to be equivalent to Eq.~\eqref{eq:thetaFromTotalTorsion} if one rewrites the derivatives of the Cartesian coordinates in the latter in terms of $\phi$ and $\chi$ using Eq.~\eqref{eq:rdot}. We also note that since the Hamiltonian only depends on local properties of the curve (namely its curvature and torsion), it follows that the corresponding evolution operator will remain invariant under rigid rotations and translations of the curve. Thus, it is really the orientation of the final tangent vector {\it relative} to the initial one that determines the final evolution operator (along with the total torsion), and not its orientation with respect to fixed coordinate axes.}

As a first example of how this geometrical structure can be exploited to design DCGs, let's take the target gate operation to be one of the Clifford gates: $U_0(T)=R(-\hat x+\hat y+\hat z,2\pi/3)$, i.e., a rotation about the axis $-\hat x+\hat y+\hat z$ by angle $2\pi/3$. To obtain a pulse that generates this gate while canceling first-order errors, we construct a closed curve that has the appropriate slope as it returns to the origin, as shown in Fig.~\ref{fig:curve1}(a). The control fields extracted from the curvature and torsion are shown in Fig.~\ref{fig:curve1}(b). A plot of the infidelity of the resulting gate as a function of the noise strength is shown in Fig.~\ref{fig:curve1}(c), where for comparison, we also show the result for a square pulse of the same duration. {The infidelity is defined in accordance with Ref.~\cite{bowdrey2002fidelity}.} It is evident that the noise-suppressing pulse makes the operation orders of magnitude more robust than a naive square pulse, and the slope of the log-log infidelity plot shows that indeed the first-order error is cancelled.

\begin{figure}
 \centering
\includegraphics[width=0.7\columnwidth]{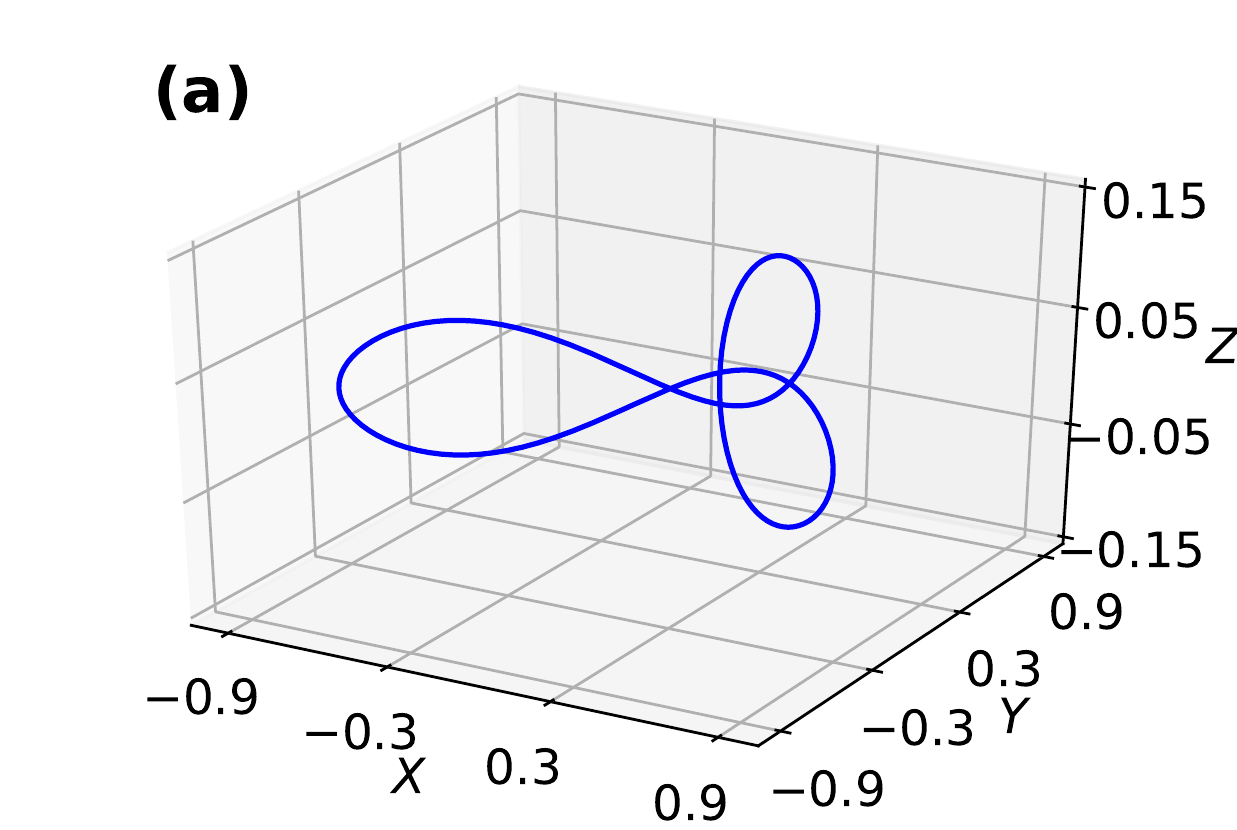}
\includegraphics[width=0.7\columnwidth]{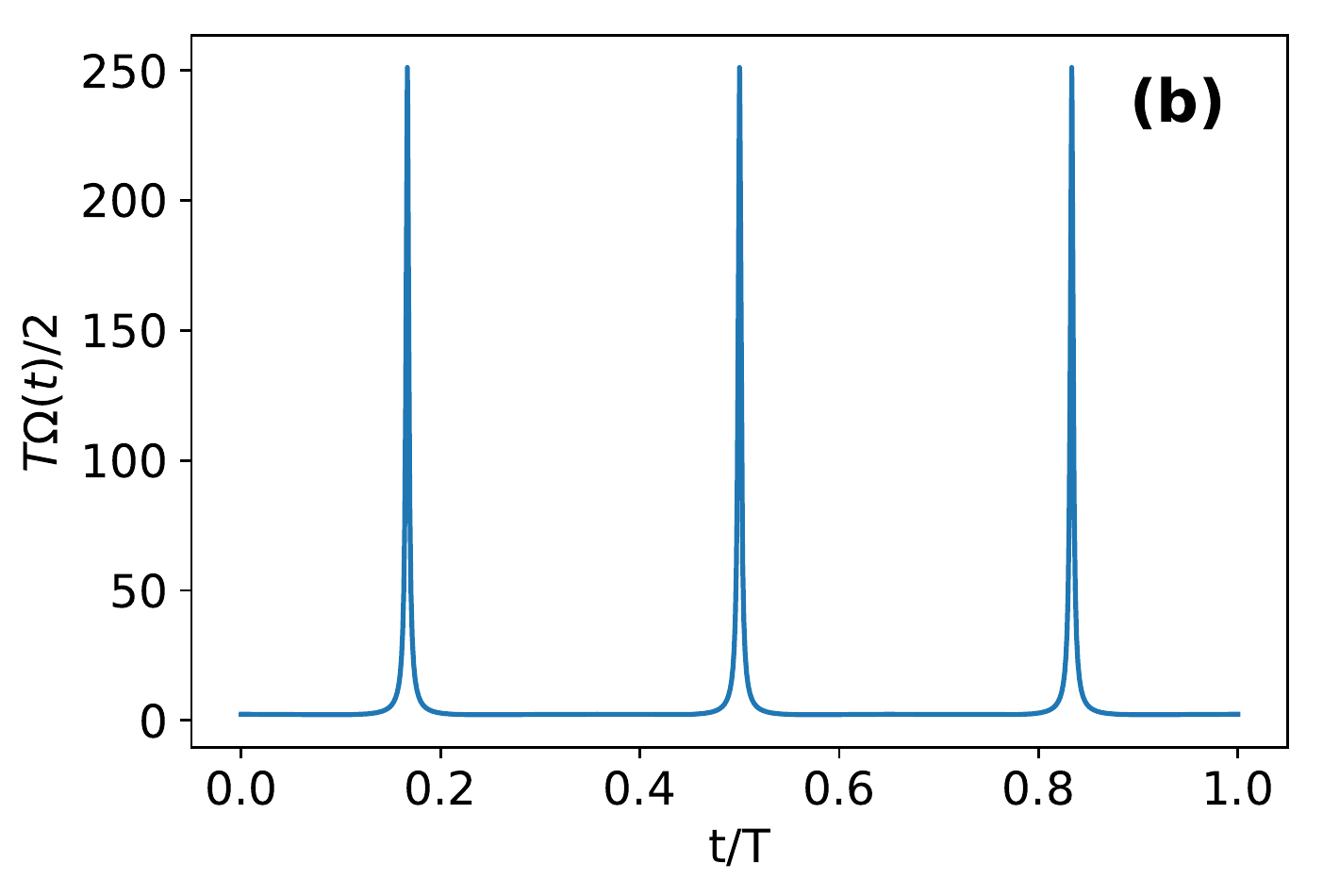}
\caption{Single-qubit identity gate for Hamiltonian $\mathcal{H}(t)=\Omega(t)\sigma_x+(\Delta+\delta\beta)\sigma_z$ that is robust to first order.}
\label{fig:curve2}
\end{figure}

In many experimental setups, there is only one control field in the Hamiltonian, say along $\sigma_x$, while there is a constant detuning or drift parameter which is noisy, so that the Hamiltonian has the form $\mathcal{H}(t)=\Omega(t)\sigma_x+(\Delta+\delta\beta)\sigma_z$. In this case, if we want to cancel the noise errors, we need to find closed curves that have constant torsion. The search for such curves is an active research area in differential geometry \cite{weiner1977closed,calini1998backlund,ivey2000minimal,bates2013curves}. Here, we provide a simple, explicit example of a robust identity operation obtained from such a curve using the recipe provided in Ref.~\cite{weiner1977closed}. Let $\alpha$ be a closed curve lying on a unit sphere which can be parameterized as $\mathbf{\alpha}(\lambda)=(x_\alpha(\lambda),y_\alpha(\lambda),z_\alpha(\lambda))$, where
\begin{equation}
  \begin{split}
  x_\alpha(\lambda)&=\frac{1}{4} \left(\sqrt{2} \cos (2 \lambda )-2 \cos (\lambda )\right),\\
  y_\alpha(\lambda)&=\frac{1}{4} \left(-\sqrt{2}\sin (2 \lambda )-2\sin (\lambda )\right),\\
  z_\alpha(\lambda)&=\frac{1}{2} \sqrt{\sqrt{2} \cos (3 \lambda )+\frac{5}{2}},\label{eq:alpha}
  \end{split}
\end{equation}
and where $\lambda \in[0,2\pi)$. A closed curve with constant torsion is given by $\gamma(\lambda)=\int \mathbf{\alpha}(\mu) \times \mathbf{\alpha}'(\mu)d\mu$. The space curve parameterized by $\gamma$ and its associated pulse shape are shown in Fig.~\ref{fig:curve2}. It is worth mentioning that the curve $\alpha$ is an example of the spherical curve formulation introduced in \cite{Barnes_SciRep15} to treat the DCG problem for Hamiltonians with a constant detuning parameter. More generally, the curves in that formulation correspond to what is known as the binormal indicatrix of a three-dimensional space curve $\gamma$. Thus, the two formulations are equivalent in the case of constant torsion, although unlike the methods of \cite{Barnes_SciRep15}, the present space curve approach provides a simple geometrical interpretation of the error-cancellation condition, and it works not only for first-order cancellation but also second-order, as we explain next.
 
\begin{figure}[ht]
\centering
\includegraphics[width=0.7\columnwidth]{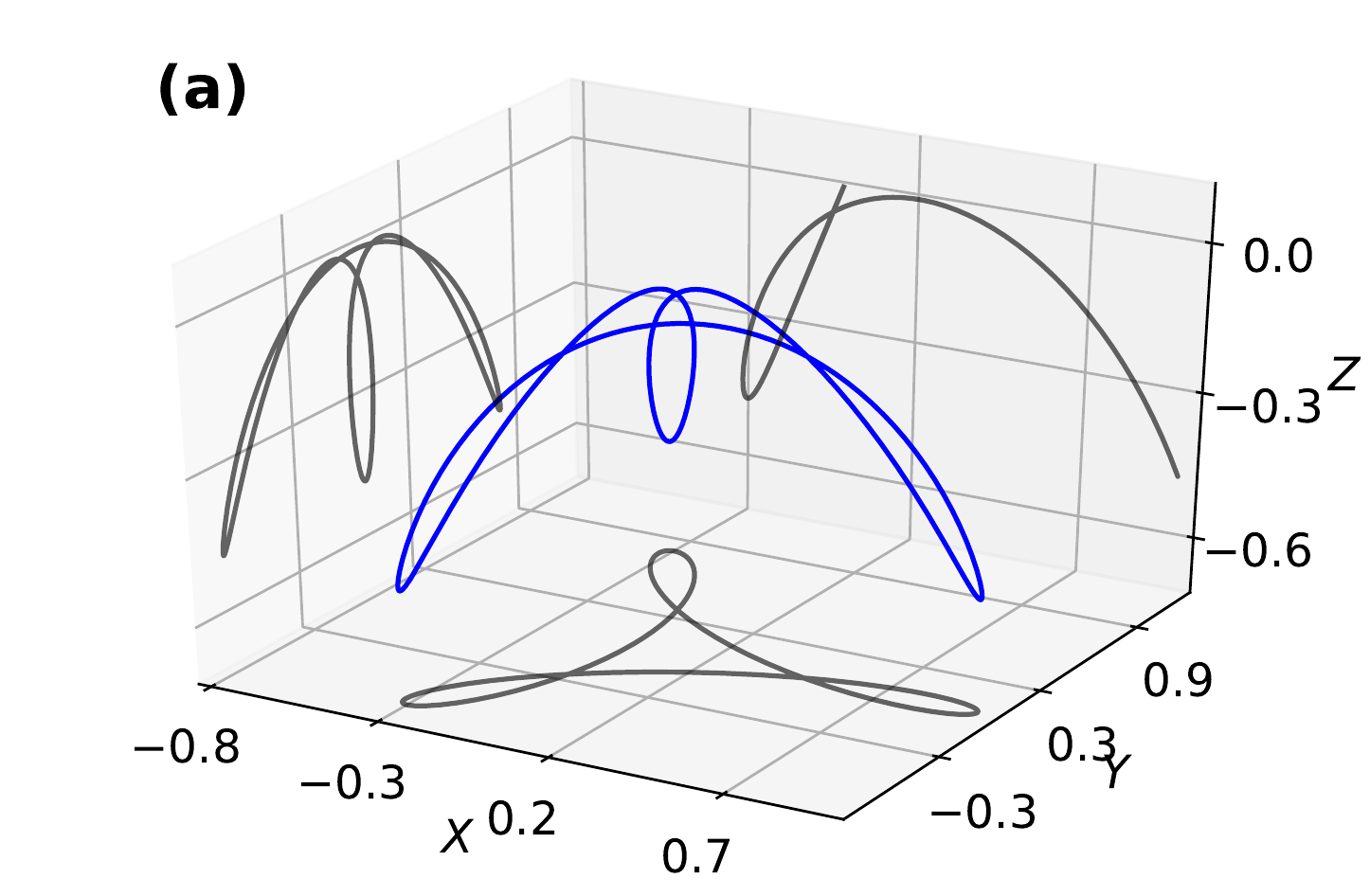}
\includegraphics[width=0.7\columnwidth]{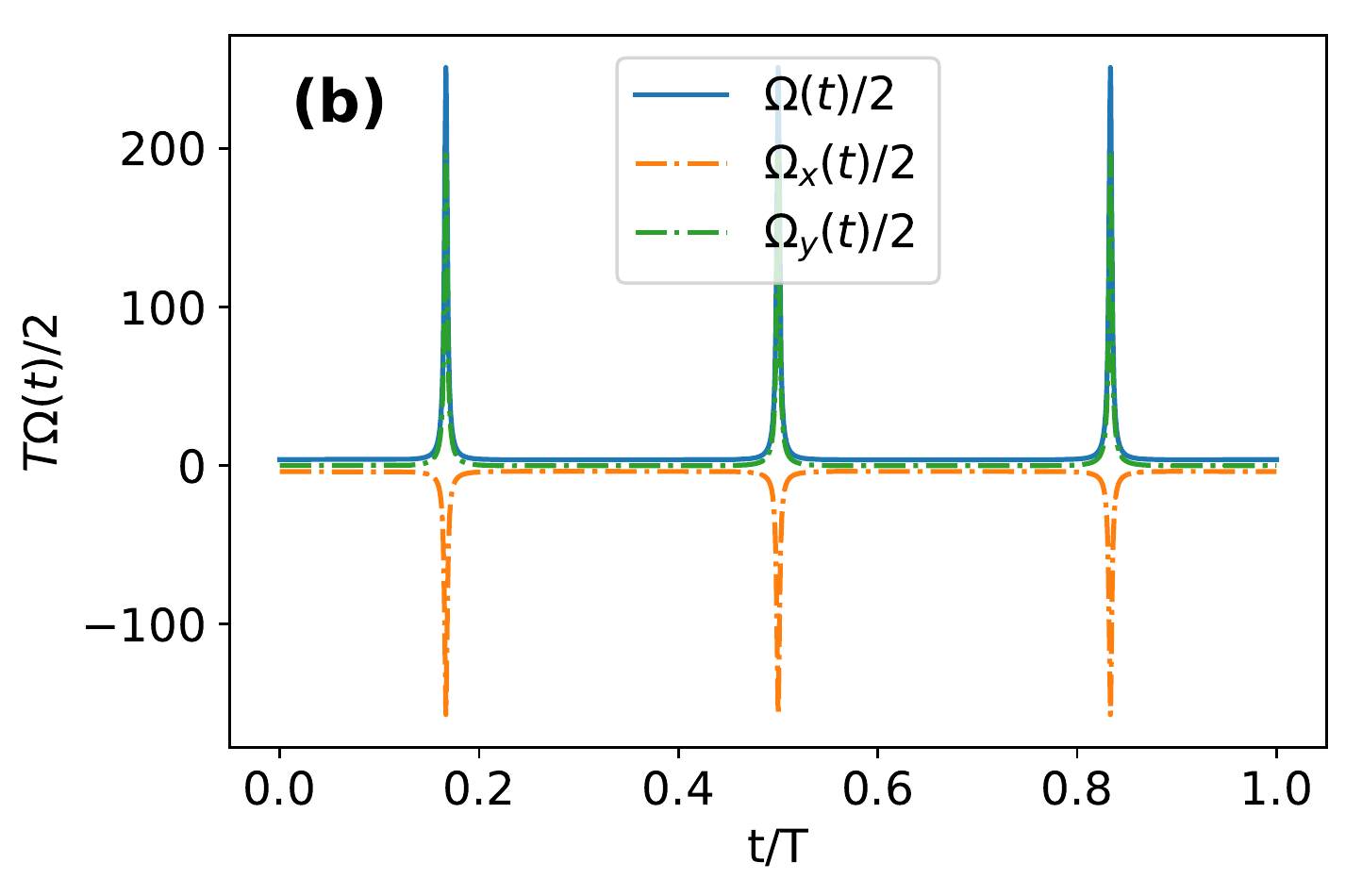}
\caption{Single-qubit identity gate robust against errors up to second order. (a) The curve () and its projections onto the $xy$, $yz$ and $xz$ planes (gray). All three projected curves have zero enclosed area. (b) The pulses obtained from the curvature and torsion of the curve in (a). Here, $\Omega_x=\Omega\cos\Phi$, $\Omega_y=\Omega\sin\Phi$.}
\label{fig:curve3}
\end{figure}

\begin{figure*}
\centering
\includegraphics[width=1.9\columnwidth]{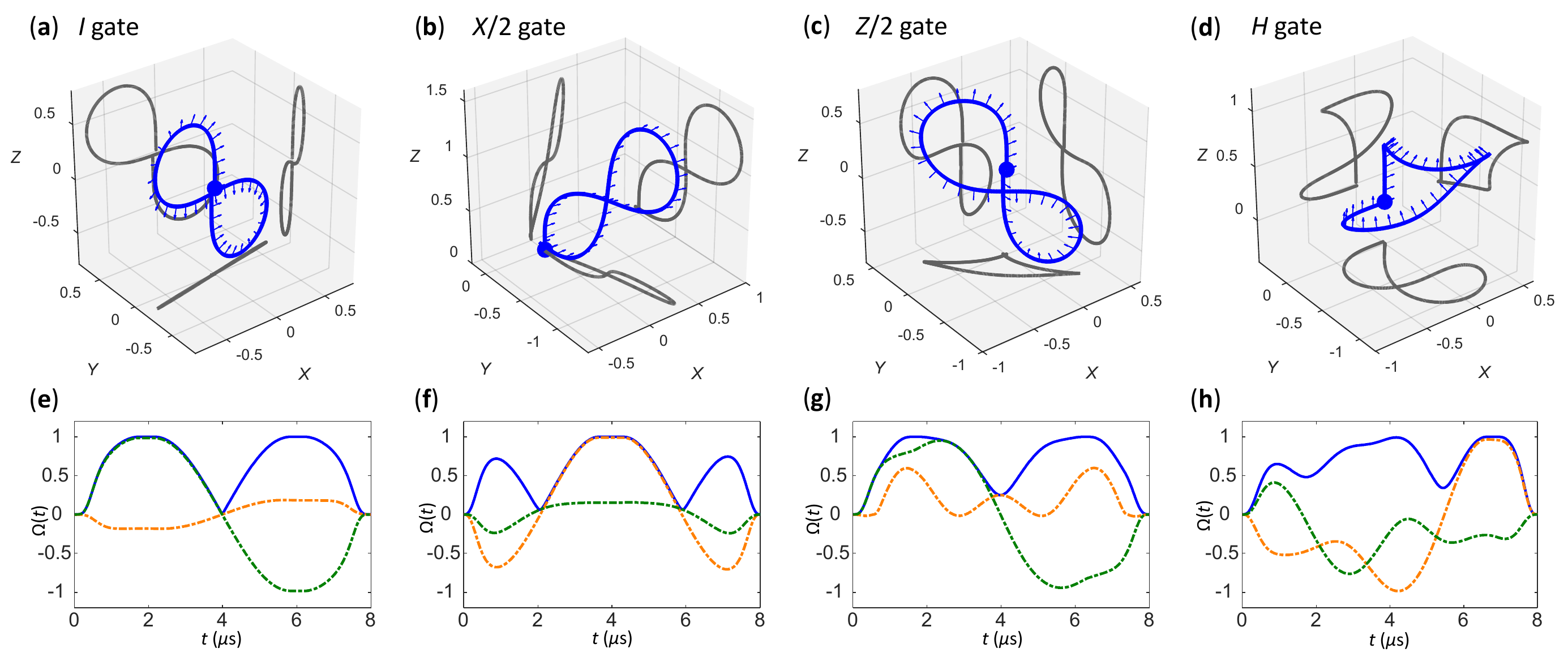}
\caption{Using space curves to analyze pulses obtained from GRAPE. (a-d) The space curves and their projections onto the $xy$, $yz$, and $xz$ planes corresponding to four different microwave pulses (e-h) designed to implement four different single-qubit gates (identity, $\pi/2$ rotation about $x$ ($X/2$), $\pi/2$ rotation about $z$ ($Z/2$), and Hadamard gate ($H$)) while cancelling noise in a silicon quantum dot spin qubit \cite{yang2018optimization}. Arrows on the curves represent the phase $\theta(t)$ in the evolution operator. For example, the $I$ gate in (a) has $\theta(T)=0$, while the $Z/2$ gate in (c) has $\theta(T)=\pi/2$. In panels (e-h), the dashed orange and green curves are $\Omega_x(t)/2$ and $\Omega_y(t)/2$, while the solid  curve is the total magnitude of the pulse envelope, as in previous figures.}
\label{fig:grapes}
\end{figure*}

We now show that second-order DCGs correspond to closed curves with vanishing-area planar projections. For second-order error cancellation, we need to impose $A_2(T)=0$. To find the curves (and hence pulses) that achieve this, we first rewrite $A_2$ as $A_2(t)=-i \mathbf{R}_2(t)\cdot\hat{\sigma}$, where $\mathbf{R}_2(t)=\int_0^t\mathbf{r}(t_1)\times\dot{\mathbf{r}}(t_1)d t_1$, as readily follows from Eqs.~\eqref{eq:errcancellation} and \eqref{eq:A1}. The constraint on error cancellation then becomes $\mathbf{R}_2(T)=(R_{2x}(T),R_{2y}(T),R_{2z}(T))=0$. When the first-order error-cancellation constraint is satisfied ($\mathbf{r}(T)=0$), $R_{2x}(T)$, $R_{2y}(T)$ and $R_{2z}(T)$ are proportional to the areas enclosed by the closed curve projected onto the $yz$, $zx$ and $xy$ planes. The sign of the area is determined by the direction of the winding of the curve.

Noticing that the curve $\mathbf{\alpha}$ defined in Eq.~\eqref{eq:alpha} already satisfies the constraint $\mathbf{R}_2(T)=0$, we can use this curve itself to generate an example of a driving pulse that cancels second-order error. The curve $\alpha$ and its projections onto the $xy$, $xz$ and $yz$ planes are shown in Fig.~\ref{fig:curve3}(a). All three projected plane curves have zero enclosed area. The pulse shape extracted from this curve (Fig.~\ref{fig:curve3}(b)) performs a robust identity operation.

In addition to facilitating the design of globally optimal control pulses, our geometrical framework can also be used to extract information about the noise-cancellation properties of pulses obtained via other means, for example using numerical algorithms such as Gradient Ascent Pulse Engineering (GRAPE) \cite{Khaneja_2005}. To exemplify this, we analyze pulses that were recently designed to implement high-fidelity single-qubit gates on silicon quantum dot spin qubits using GRAPE \cite{yang2018optimization}.
Fig.~\ref{fig:grapes} shows the space curves for four such pulses, which perform four different single-qubit gates, including an identity operation ($I$), a $\pi/2$ rotation about $x$ ($X/2$), a $\pi/2$ rotation about $z$ ($Z/2$), and a Hadamard operation ($H$). The GRAPE algorithm is implemented with gate fidelity as the cost function and with a noise level corresponding to $\sqrt{\left<\delta\beta^2\right>}=16.7$ kHz, which was attributed to nuclear spin noise in \cite{yang2018optimization}. Constraints are also imposed on the pulse bandwidth through filtering, where the pulses are strongly smoothed out and forced to approach zero at the beginning and end of the gate. We have included arrows along the space curves to indicate the value of the evolution operator phase $\theta(t)$ as the system evolves. The value of this phase at the final time, $\theta(T)$ distinguishes between some of the gates, for example the $I$ and $Z/2$ gates. Note that if we were to include a third driving field, $\Omega_z(t)$, then this would provide direct control over the phase $\theta(t)$.

From the figure, it is evident that in each case, the corresponding space curve is (almost) closed, showing that the first-order error-cancellation constraint is almost perfectly satisfied. Moreover, the two-dimensional projections of the curves form figure-eight shapes in most cases, showing that the second-order cancellation constraint is nearly satisfied as well. Interestingly, it was found that these pulses needed to be 4-5 times longer than the typical time scale of a $\pi$ pulse (1.75 $\mu$s for the parameters used in Ref.~\cite{yang2018optimization}); the reason for this is apparent from the space curve, where the bandwidth constraints require pulse durations on the order of 8 $\mu$s in order for the planar projections of the curves to complete their respective figure-eights and thus suppress second-order noise. It is clear from these results that experimental limitations on pulse amplitude or bandwidth are fully compatible with the space curve formalism, and that realistic pulses correspond to smooth curves that respect the geometrical noise-cancellation conditions.

While here we have focused on two-dimensional Hilbert spaces, the idea can in principle be generalized to higher-dimensional systems. This can be done by decomposing $A_1(t)$ into tensor products of Pauli matrices. For example, for a two-qubit system, this will lead to a mapping between robust pulses and closed curves in a 15-dimensional space. The higher-dimensional form of the Frenet-Serret equations can relate the generalizations of curvature and torsion for these curves to driving fields in the two-qubit Hamiltonian.

In conclusion, we uncovered a general geometrical framework hidden within the Schr\"odinger equation that yields the entire solution space of pulses that implement dynamically corrected single-qubit gates in the presence of quasistatic noise. Pulses that cancel first-order noise errors can be obtained from closed space curves in three dimensions, while curves that have the additional property that their planar projections have vanishing enclosed area guarantee the cancellation of second-order errors as well. We demonstrated these findings with explicit examples of closed curves and the pulses they correspond to and showed that a similar framework holds for higher-dimensional Hilbert spaces as well. Our findings open up the possibility of obtaining globally optimal control fields for a wide range of physical systems and types of noise.

E.B. would like to thank Renbao Liu for helpful comments. E.B. acknowledges support from the U.S. Army Research Office
(Grant No.W911NF-17-0287) and from the U.S. Office of Naval Research (Grant No. N00014-17-1-2971). A.D. and C.H.Y. acknowledge support from U.S Army Research Office (Grant No. W911NF-17-1-0198).

\bibliographystyle{apsrev}

\bibliography{note}
\end{document}